\documentclass[3p, sort&compress]{elsarticle} 

\usepackage{graphicx}
\usepackage{dcolumn}
\usepackage{bm}

\usepackage{epstopdf}
\newcommand{\Si}{Si(111)--(1$\times$1) }
\newcommand{\Sii}{Si(111)--(7$\times$7) }

\newcommand{\TeSii}{Te/Si(111)--($2\sqrt{3}\times2\sqrt{3}$)R$30^{\circ}$ }

\usepackage{lineno,hyperref}
\usepackage{url} 

\begin{document}
	
	\begin{frontmatter}
	
		\title{Surface structures of tellurium on Si(111)--(7$\times$7) studied by low-energy electron diffraction and scanning tunneling microscopy}

		\author[address1,address2]{Felix L\"upke\corref{presentaddress}}

		\cortext[presentaddress]{Present address: Department of Physics, Carnegie Mellon University, Pittsburgh, PA 15213, USA}
		
		\author[address3]{Ji\v{r}\'i Dole\v{z}al}
		
		\author[address1,address2]{Vasily Cherepanov}
		
		\author[address3]{Ivan O\v{s}t'\'adal}
		
		\author[address1,address2]{F. Stefan Tautz}
		
		\author[address1,address2]{Bert Voigtl\"ander\corref{mycorrespondingauthor}}
		\ead{b.voigtlaender@fz-juelich.de}
		
		\cortext[mycorrespondingauthor]{Corresponding author}
		\address[address1]{Peter Gr\"unberg Institut (PGI-3), Forschungszentrum J\"ulich, 52425 J\"ulich, Germany}

		\address[address2]{J\"ulich Aachen Research Alliance (JARA), Fundamentals of Future Information Technology, 52425 J\"ulich, Germany}
		
		\address[address3]{Department of Surface and Plasma Science, Faculty of Mathematics and Physics, Charles University, 182 00 Prague 8, Czech Republic}
		
		\begin{abstract}
			The Te-covered Si(111) surface has received recent interest as a template for the epitaxy of van der Waals (vdW) materials, e.g. Bi$_2$Te$_3$. Here, we report the formation of a Te buffer layer on \Sii by low-energy electron diffraction (LEED) and scanning tunneling microscopy (STM). While deposition of several monolayer (ML) of Te on the \Sii surface at room temperature results in an amorphous Te layer, increasing the substrate temperature to $770\rm\,K$ results in a weak (7$\times$7) electron diffraction pattern. Scanning tunneling microscopy of this surface shows remaining corner holes from the \Sii surface reconstruction and clusters in the faulted and unfaulted halves of the (7$\times$7) unit cells. Increasing the substrate temperature further to $920\rm\,K$ leads to a \TeSii surface reconstruction. We find that this surface configuration has an atomically flat structure with threefold symmetry.
		\end{abstract}
		
		
	\end{frontmatter}
	

\section{Introduction}
In recent years, layered vdW materials such as ZnTe, FeTe and topological insulators Bi$_2$Te$_3$, Sb$_2$Te$_3$ as well as ternary and quaternary compounds thereof have received increasing interest. Their unique electronic properties and topologically non-trivial phases make them promising candidates for an application in future electronic devices \cite{Hsieh2009,Roushan2009,Mak2010, RadisavljevicB.2011, Luepke2017,Luepke2018}. In this context, passivated semiconductor surfaces have gained importance as substrates for the growth of the layered materials by vdW epitaxy \cite{Luepke2017a}. In this growth-mode, an initial passivation of the substrate surface is required to grow high-quality films, ideally by one of the chemical elements of the subsequently grown films. The purpose of the buffer layer is to saturate any dangling bonds of the substrate and to form an atomically flat template for the film growth. Furthermore, the buffer layer needs to be insulating to not result in parallel conduction channels which undermine the applicability of the on top grown films in devices \cite{Luepke2017a}. Due to the typically hexagonal crystal structure of the tellurium-based vdW materials, Si(111) has emerged as one of the most frequently used substrates. While tellurium is known to saturate dangling bonds of Si(100) surfaces, as used for instance in surfactant-mediated growth of germanium \cite{Bennett1997}, there is only a very limited amount of studies on the Te growth on Si(111) surfaces \cite{Shih1976, Kanai2006}, all in the sub-monolayer coverage regime. In the present study, we report the growth of Te on \Sii by a combined low-energy electron diffraction (LEED) and scanning tunneling microscopy (STM) study.

\section{Methods}
\begin{figure*}
	\centering
	\includegraphics[width=16cm]{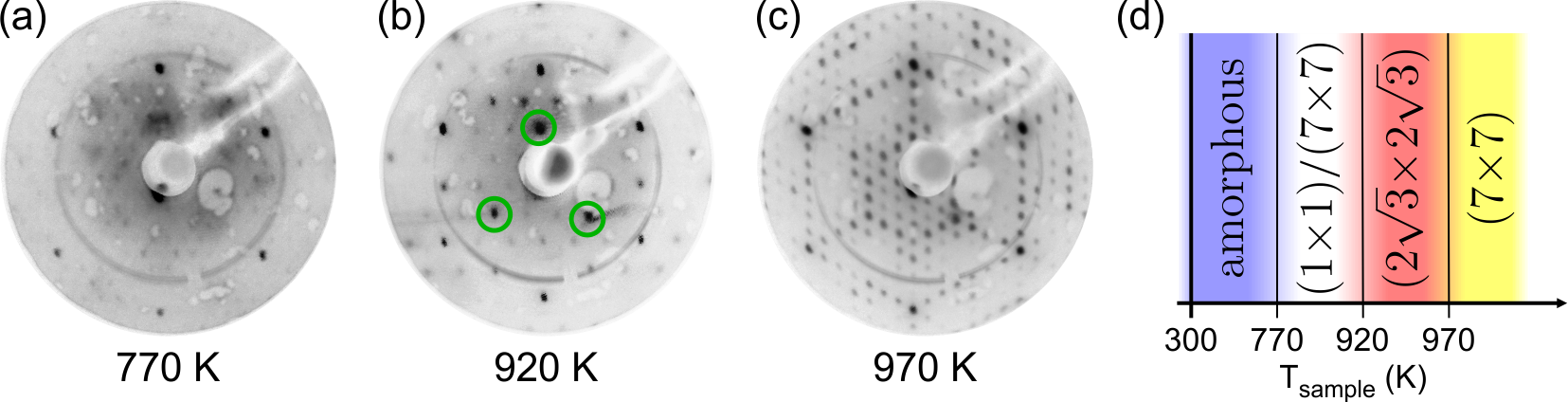}
	\caption{\label{Overview} (color online) LEED pattern of $10\rm\,ML$ Te deposited on \Sii at room-temperature and subsequently annealed to the indicated temperatures. (a) LEED pattern after annealing at $770\rm\,K$ showing mainly a (1$\times$1) periodicity with additional weak (7$\times$7) spots with respect to the Si(111) substrate. (b) LEED pattern after annealing at $920\rm\,K$ showing a ($2\sqrt{3}\times2\sqrt{3}$)R$30^{\circ}$ periodicity which is rotated by $30^{\circ}$ with respect to the Si(111) spots. The inner spots of the pattern show an increased intensity corresponding to a threefold symmetry, as indicated by the green circles. (c) After heating the sample above 970$\,\rm K$ the \Sii reconstruction reappears. (d) Schematic of the Te/Si(111) surface phases as a function of sample temperature.}
\end{figure*}
The Si(111) sample was cleaned under ultra-high vacuum conditions (base pressure below $1\times10^{-10}\rm\,mbar$) by repeated cycles of annealing at $1500\,\rm K$ for $30\,\rm s$. The sample temperature $T_{\rm sample}$ was determined using an infrared pyrometer. After the last annealing cycle, the sample was quenched from $1500\,\rm K$ to $1170\,\rm K$ to avoid step bunching \cite{Latyshev1989}, followed by a slow cooling ($1\,\rm K/s$) to $1070\,\rm K$ to form a long-range ordered \Sii reconstruction. Subsequently, the sample was cooled down to RT by radiative cooling for $1\rm\, h$ and its cleanliness was confirmed by LEED and STM. Te was deposited onto the \Sii substrate from a Knudsen cell, the deposition rate of which was calibrated by a quartz crystal microbalance. The home-built room-temperature Besocke-type STM was equipped with electrochemically etched tungsten tips.

In the following, two separate experiments were performed: 1) Solid phase epitaxy, in which Te was deposited onto the \Sii surface at room temperature. The sample was then gradually heated under continuous LEED observation. 2) Epitaxial growth of Te on the \Sii surface at elevated temperatures, followed by cooling the sample to room temperature and subsequent LEED and STM analysis.

\section{Results and Discussion}

\subsection{Solid phase epitaxy under LEED observation}
$10\rm\, ML$ of Te was deposited onto the \Sii surface at a sample temperature $T_{\rm sample}=300\,\rm K$ ($1\,\rm ML$ corresponds to $7.84\times10^{14}\,\rm atoms/cm^2$, i.e. half a Si(111) bilayer). Next, the sample was placed in the rear view LEED setup and the temperature of the sample was increased at a rate of $10\rm\,K/min$. This measurement allows to directly observe changes in the surface structure as a function of the sample temperature. The LEED electron energy in this experiment was fixed at $35\rm\,eV$. To determine the temperature of the sample in the required temperature range which was partially below the minimum specification of the infrared pyrometer ($530\rm\,K$), an interpolation of the temperature vs. heating current calibration curve was used; measured at high temperatures using the pyrometer and going through room temperature at zero heating current. The resulting accuracy of the temperature measurement was estimated to be $\pm20\rm\,K$.

Starting from room temperature, only a diffuse background was observed on the LEED screen, indicating an amorphous state of the deposited Te. Once the sample reached $T_{\rm sample}=770\rm\,K$, a (1$\times$1) pattern corresponding to the periodicity of the Si(111) substrate is observed (Fig. \ref{Overview} (a)). In addition to the clear (1$\times$1) spots, there is a weak (7$\times$7) pattern and still a relatively high background signal. This observation is explained by the bulk of the amorphous Te layer desorbing from the \Sii surface, leaving behind only a fraction of the deposited Te. The high background signal and weak (7$\times$7) pattern are an indication of inhomogeneities in the surface layer. The (1$\times$1)/(7$\times$7) periodicity remains upon further increasing the sample temperature up to $T_{\rm sample}=920\rm\,K$. At this temperature, there is a change in the LEED pattern to a ($2\sqrt{3}\times2\sqrt{3}$)R$30^{\circ}$ periodicity (Fig. \ref{Overview} (b)). This pattern shows sharper spots and less background signal compared to Fig. \ref{Overview} (a) and persists upon increasing the temperature up to $T_{\rm sample}=970\rm\,K$. Heating to this temperature changes the pattern to a (7$\times$7) structure resembling that of clean Si(111)--(7$\times$7) (Fig. \ref{Overview} (c)). This indicates a complete desorption of Te from the surface and restoration of the clean \Sii surface reconstruction. A schematic of the observed Te/Si(111) LEED pattern as a function of temperature is shown in Fig. \ref{Overview} (d).

\subsection{Te growth at elevated temperatures}
\begin{figure*}
	\centering
	\includegraphics{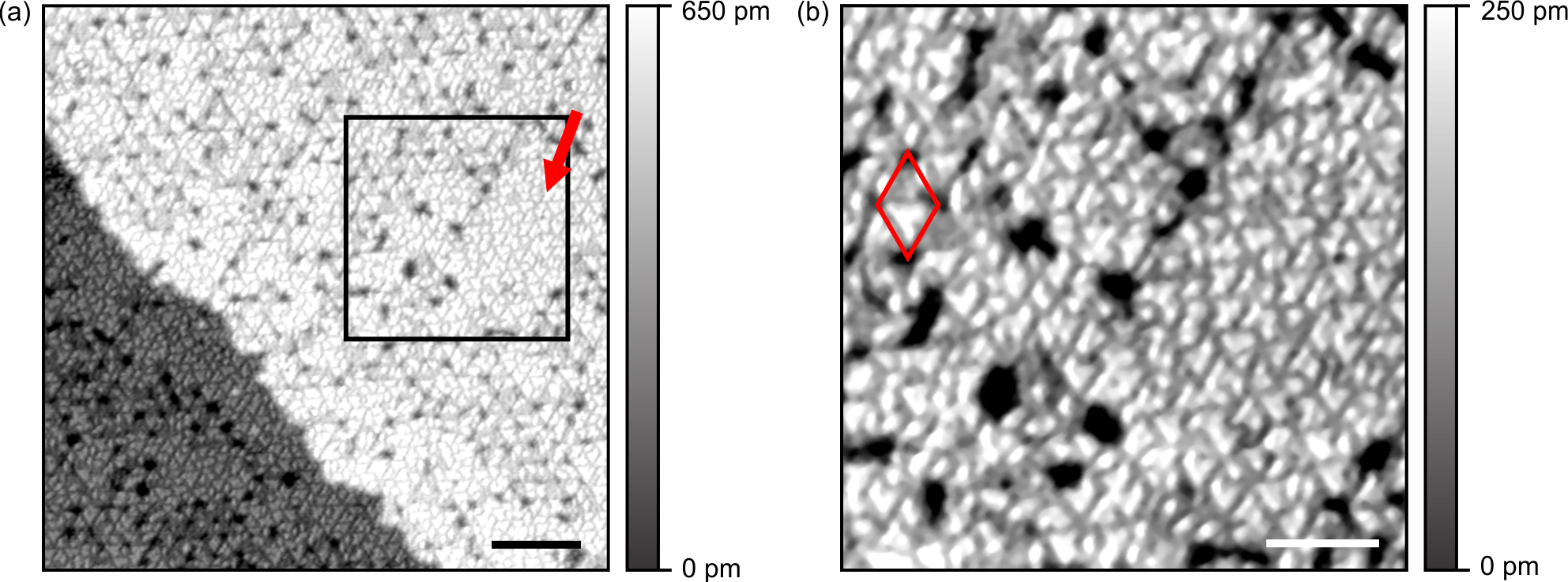}
	\caption{\label{1x1} (color online) Sample surface after deposition of $1\rm\,ML$ Te at a substrate temperature of $T_{\rm sample}=777\,\rm K$. (a) STM topography of the sample surface showing remaining corner-holes and the formation of clusters on the sample surface. In some areas, a closed surface is observed, with no remaining corner-holes (indicated by the red arrow). Scan parameters: $V_{\rm sample}=2.1\rm\,V$ and $I_{\rm t}=300\rm\,pA$. Scale bar: $10\rm\,nm$. (b) Zoom into the square indicated in (a) showing the clusters within the faulted and unfaulted unit cells of the former \Sii reconstruction (($7\times7$) unit cell indicated in red). Scale bar: $5\rm\,nm$.}
\end{figure*}

Growth of Te on \Sii at elevated substrate temperatures results in a similar LEED pattern as for the corresponding heating temperature during the solid phase epitaxy described above. The Te is deposited under direct pyrometer monitoring of the sample temperature, such that we estimate the accuracy of the sample temperature in this experiment to be $\pm10\rm\,K$. Deposition of $1\rm\,ML$ Te at a sample temperature of $T_{\rm sample}=777\,\rm K$ leads to a LEED pattern similar to the one shown in Fig. \ref{Overview} (a) with the STM measurement of this sample surface shown in Fig. \ref{1x1}. Figure \ref{1x1} (a) displays that the surface structure is dominated by clusters and periodic holes into the next lower atomic layer. The average periodicity of the holes corresponds to that of the corner-holes of the \Sii reconstruction. A smaller area scan of the sample surface (Fig. \ref{1x1} (b)) shows that clusters are located inside the faulted and unfaulted half unit cells (HUC) of the ($7\times7$) reconstruction. An explanation of this finding is that the Te sticks to the sample surface but the substrate temperature during the deposition is not high enough to lift the \Sii reconstruction. This mechanism of growth also occurs when depositing other elements on \Sii \cite{Asaoka2005,Ostadal2005,Byun2008} and is consistent with an earlier study of the initial adsorption of Te on \Sii \cite{Kanai2006}. We conclude that the clusters consist of accumulated Te atoms. The structure observed in STM is in correspondence with the LEED image showing faint (7$\times$7) spots which can be explained to stem from the remaining (7$\times$7) reconstruction which we observed in most surface areas. The (1$\times$1) LEED spots most probably stem from the Si(111) substrate below the surface layer. Variation of the substrate temperature in the given temperature range of the (1$\times$1)/(7$\times$7) structure, as well as the variation of the amount of deposited Te ($1-10\rm\,ML$) and holding the sample at elevated temperature after closing the evaporator, did not result in a uniformly closed film with no corner holes. We conclude that it is energetically unfavorable to form a closed Te surface layer below $T_{\rm sample}=920\,\rm K$.

\begin{figure*}
	\centering
	\includegraphics{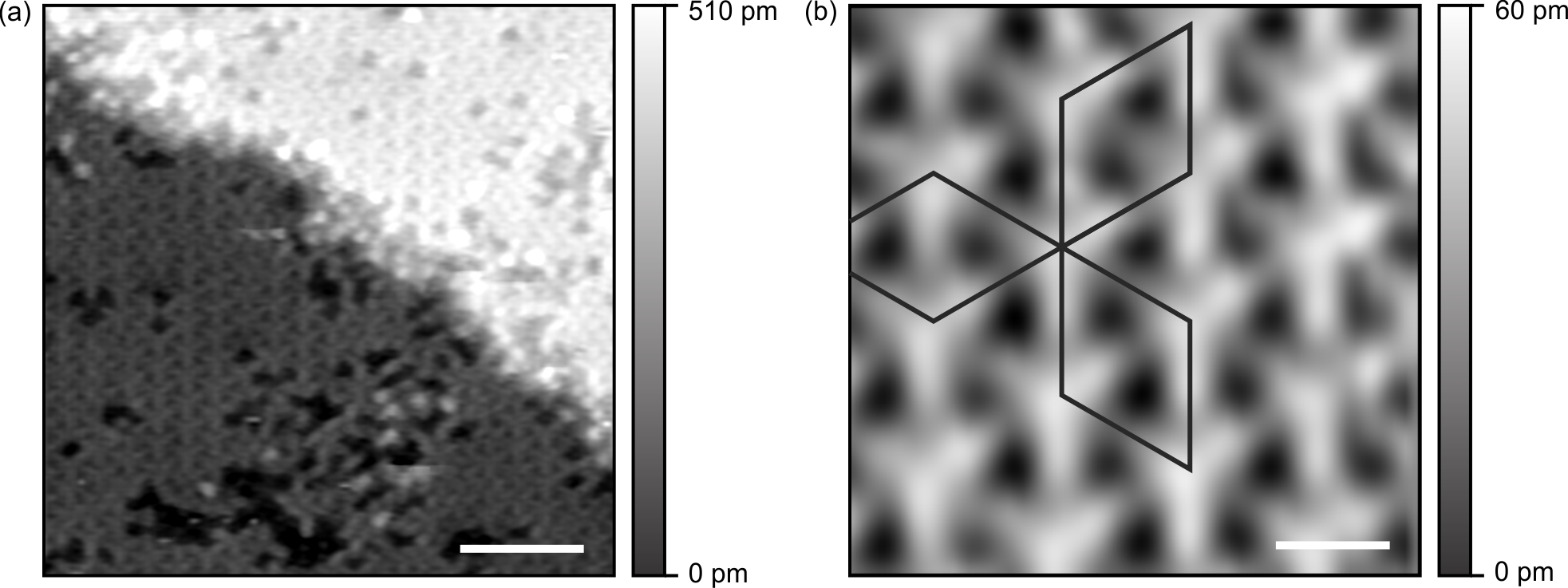}
	\caption{\label{2sqrt3} Structure of the \TeSii reconstruction. (a) STM topography showing atomically flat terraces with vacancy defects and occasional adatoms on top. Scan parameters: $V_{\rm sample}=1.5\rm\,V$ and $I_{\rm t}=15\rm\,pA$. Scale bar: $5\rm\,nm$. (b) High resolution image of the atomic structure. The three black diamonds show the threefold-symmetric \TeSii unit cell, with a lattice constant of $13.3\,$\AA. Scale bar: $1\rm\,nm$.}
\end{figure*}
Deposition of $1\rm\,ML$ Te at a sample temperature of $T_{\rm sample}=940\,\rm K$ leads to a sharp ($2\sqrt{3}\times2\sqrt{3}$)R$30^{\circ}$ LEED pattern (Fig. \ref{Overview} (c)). At this temperature, the thermal energy is high enough to lift the \Sii reconstruction during the Te deposition and the Te layer is formed directly on the \Si surface. This periodicity has been reported earlier in electron diffraction experiments \cite{Shih1976}, but to our knowledge no STM measurements were reported yet. Figure \ref{2sqrt3} (a) shows the STM topography of the sample surface. We observe atomically flat terraces with atomic-scale vacancies and isolated adsorbates on top. The observed step heights corresponds to that of one Si(111) bilayer. An atomic-scale STM image of the \TeSii reconstruction is shown in Fig. \ref{2sqrt3} (b). A threefold surface symmetry is observed with three equivalent orientations of the unit cell indicated by black diamonds, respectively (side length of $13.3\,$\AA). This result is in agreement with the LEED pattern of this surface in which three of the six inner spots of the pattern show an increased intensity (green circles in Fig. \ref{Overview} (b)). Furthermore, the \TeSii reconstruction is observed to have a strong preferential orientation, i.e. no domains of a different rotational orientation were observed in the STM measurements, which is explained to stem from the underlying threefold-symmetric \Si surface. The threefold features evident in the STM images are most likely Te oligomers, similar to the features seen in the $\beta$-Bi/Si(111)--($\sqrt3\times\sqrt3$)R$30^{\circ}$ reconstruction \cite{Shioda1993, Kuzumaki2010}.

\section{Conclusion}
In summary, we find that deposition of several ML of Te on \Sii at room temperature results in an amorphous Te layer. In the temperature range $T_{\rm sample}=770\rm\,K-920\rm\,K$ Te forms a Te/Si(111) surface phase with weak (7$\times$7) LEED pattern, in which the Te forms clusters in the faulted and unfaulted halves of the \Sii unit cell. In the temperature range $T_{\rm sample}=920\rm\,K-970\rm\,K$ Te forms a \TeSii phase for which we report an atomically flat surface structure with threefold symmetry. The clean \Sii structure can be recovered by heating to $T_{\rm sample}=970\rm\,K$. As mentioned above, for the application of devices grown on a Te buffer layer, it is important to check the conductivity of the Te/Si(111) buffer layer to ensure that it does not result in parallel conduction channels, undermining the applicability of the grown material. For example, the $\beta$-Bi/Si(111)--($\sqrt3\times\sqrt3$)R$30^{\circ}$ is unsuitable as a substrate for devices due to its high electrical conductivity \cite{Just2015}. The measurement of the electrical conductivity of surface layers can be achieved e.g. by {\it in situ} four-probe measurements \cite{Luepke2018, Luepke2018a}, or scanning tunneling potentiometry \cite{Muralt1986,Luepke2015}.

\section*{Declarations of interest}
none



\end{document}